# The IFF Approach to the Lattice of Theories

> *Systems, scientific and philosophic, come and go. Each method of limited understanding is at length exhausted. In its prime, each system is a triumphant success. In its decay, it is an obstructive nuisance.*
>
> Alfred North Whitehead
>
> *Every theory [in the lattice of theories] states a possibly useful "method of limited understanding", which may be a "triumphant success" in its prime, but eventually becomes "an obstructive nuisance". The purpose of the lattice is to provide a systematic way of relating all possible ontologies in order to facilitate the inevitable upgrades and conversions.*
>
> *A framework is created which can support an open-ended number of theories (potentially infinite) organized in a lattice together with systematic metalevel techniques for moving from one to another, for testing their adequacy for any given problem, and for mixing, matching, combining, and transforming them to whatever form is appropriate for whatever problem anyone is trying to solve.*
>
> John Sowa

## Overview

The IFF approach for the notion of "lattice of theories" uses the idea of a *concept lattice* from Formal Concept Analysis (Ganter and Wille) and the idea of the *truth classification* from Information Flow (Barwise and Seligman). See the IFF references. The IFF approach is concentrated in the joining of these two important ideas. The result is called the *truth concept lattice*, the concept lattice of the truth classification. The IFF provides a principled (versus ad hoc) approach for John Sowa's "lattice of theories" framework. The "lattice of theories" is represented by the truth concept lattice, and the theories quoted above are represented by the formal concepts in the truth concept lattice. Suitable set-theoretic foundational concerns are covered in the IFF Category Theory (meta) Ontology (IFF-CAT) and the IFF Upper Classification (meta) Ontology (IFF-UCLS). The IFF approach provides a structuring methodology for the SUO metalevel.

In the following, we use the abbreviations FCA for Formal Concept Analysis and IF for Information Flow. The theory of FCA and the theory of IF have several notions in common[i]. What the FCA has, but the IF lacks, is the notion of a concept lattice[ii]. As evident from the following discussion, any classification has an associated concept lattice, whose elements are called formal concepts[iii].

## *Objects*

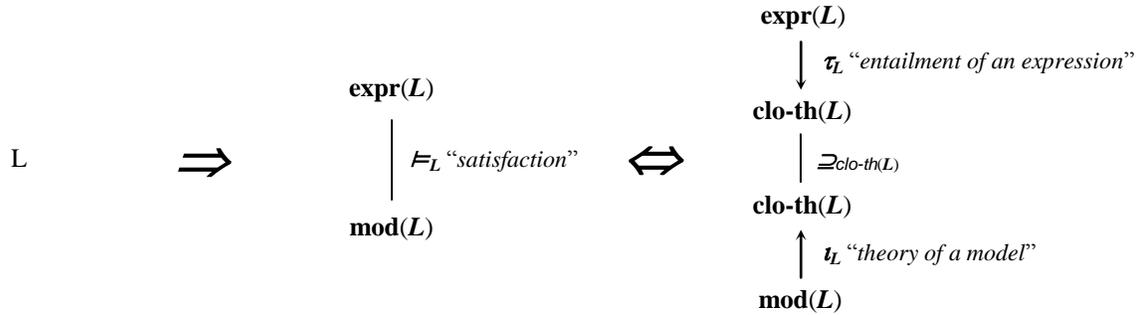

## Truth Classification

The idea of the *truth classification* is described briefly in Example 4.6 on page 71 of the text by Barwise and Seligman. We paraphrase here. Given a first order language $L$, the truth classification of $L$

$\quad$ truth-classification$(L) = \langle$expr$(L)$, mod$(L)$, $\models_L\rangle$

is the (set-theoretically large) classification whose types are expressions of $L$, whose instances are model-theoretic structures of $L$, and whose classification relation

$\quad M \models \varphi$

is satisfaction, meaning the expression $\varphi$ is true in the model-theoretic structure $M$. The type set of a model $M \in$ mod$(L)$ is the set of all $L$-expressions true in $M$, usually called *the theory of $M$*. The instance set of an expression $\varphi \in$ expr$(L)$ is the collection of all models of $\varphi$.

## Truth Concept Lattice

With these definitions out of the way, we can now define the truth concept lattice. The metatheoretic concept lattice of the truth classification is called the *metatheoretic truth concept lattice*. Given a first order language $L$, the metatheoretic truth concept lattice of $L$

$\quad \langle$conc$(L)$, expr$(L)$, mod$(L)$, $\iota_L$, $\tau_L\rangle$

is the (set-theoretically large) concept lattice whose types are expressions of $L$, and whose instances are model-theoretic structures of $L$. A formal concept $c \in$ conc$(L)$ of the truth classification of $L$ is a pair $c = ($extent$_L(c)$, intent$_L(c))$, where intent$_L(c)$ is a closed theory (set of expressions) and extent$_L(c)$ is the class of all models for that theory. The IFF axiomatic representation is an analog to the metatheoretic truth concept lattice called the *axiomatized truth concept lattice*

$\quad$ truth-concept-lattice$(L) = \langle$clo-th$(L)$, expr$(L)$, mod$(L)$, $\iota_L$, $\tau_L\rangle$

over a type language $L$. In the axiomatic representation the formal concepts in conc$(L)$ are represented by their intents in clo-th$(L)$ – the closed theories of $L$. The lattice order

$\quad c_1 \leq_L c_2$,

is the opposite of theory inclusion. The join or supremum of two theories is the intersection of the theories. The meet or infimum of two theories is the theory of the common models. Both $L$-models and $L$-expressions generate formal truth concepts (closed theories). An *object concept* is the theory of a model. An *attribute concept* is the entailment theory of an expression.

## Semantic Entailment

Semantic entailment is a binary relation between a theory and an expression. Semantic entailment is equivalent to the truth concept lattice order. For any 1$^{st}$-order language $L$, an $L$-expression $\varphi$ is in the closure of an $L$-theory $T$ iff any $L$-model that satisfies the theory also satisfies the expression. Clearly, a theory is contained in its closure. In the IFF theory namespace, a closure function maps theories to their closure. The closure function and the truth concept lattice order on closed theories induce an order on theories. For any $1^{st}$-order language $L$, an $L$-theory $T_1$ is below an $L$-theory $T_2$ when the closure of the first contains the closure of the second:

$$T_1 \leq_L T_2 \text{ iff } clo(T_1) \supseteq clo(T_2).$$

Semantic entailment is defined as the order between a theory and the singleton of an expression. For any $1^{st}$-order language $L$, an $L$-theory $T_1$ entails an $L$-expression $\varphi$ when the theory is below the singleton theory of the expression:

$$T_1 \vdash_L \varphi \text{ iff } T_1 \leq_L \{\varphi\} \text{ iff } clo(T_1) \supseteq \{\varphi\} \text{ iff } \varphi \in clo(T).$$

As a result, the theory order can be defined in terms of entailment:

$$T_1 \leq_L T_2 \text{ iff } (\forall \varphi_2 \in T_2) \; T_1 \vdash_L \varphi_2.$$

## Navigating the Lattice of Theories

This section follows the discussion in section 6.5 "Theories, Models and the World" of the book *Knowledge Representation* by John Sowa. From each theory in the lattice of theories, the partial ordering defines paths to the more generalized theories above and the more specialized theories below. Figure 2 shows four ways for moving along paths from one theory to another: contraction, expansion, revision and analogy.

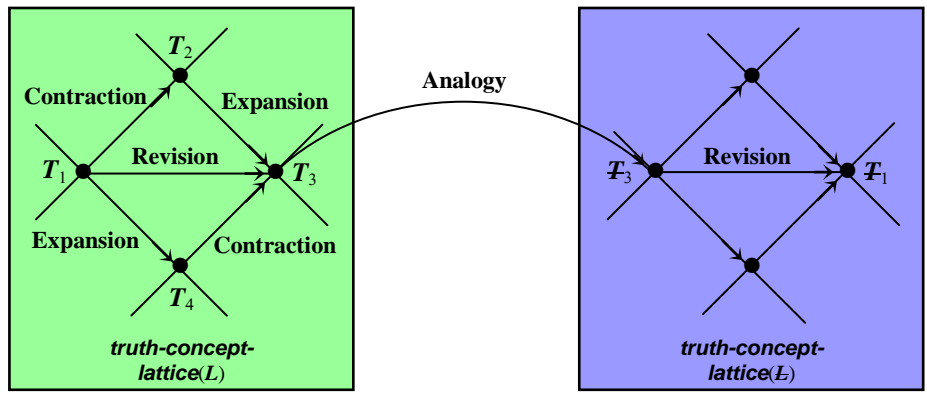

**Figure 2: Navigating the Lattice of Theories**

- **Contraction:** Any theory can be contracted or reduced to a smaller, simpler closed theory by deleting one or more axioms.
- **Expansion:** Any theory can be expanded by adding one or more axioms.
- **Revision:** A revision step is composite – it uses a contraction step to discard irrelevant details, followed by an expansion step to added new axioms.
- **Analogy:** Unlike contraction and expansion, which move to nearby theories in the lattice, analogy jumps to a remote theory by systematically renaming the entity types, relation types, and constants (individuals) that appear in the axioms.

By repeated contraction, expansion, and analogy, any theory can be converted into any other. Multiple contractions would reduce a theory to the empty theory at the top of the lattice. The top theory in the concept lattice of theories is the closure of the empty theory – it contains only tautologies or logical truths; i.e., expressions that are true in all models (it is "true of everything"). Multiple expansions would reduce a theory to the full inconsistent theory at the bottom of the lattice. The full inconsistent theory is the closed theory consisting of all expressions; i.e., expressions that are true in no models (it is "true of nothing").

The IFF represents the operations of contraction, expansion and revision in the truth namespace.

## *Morphisms*[1]

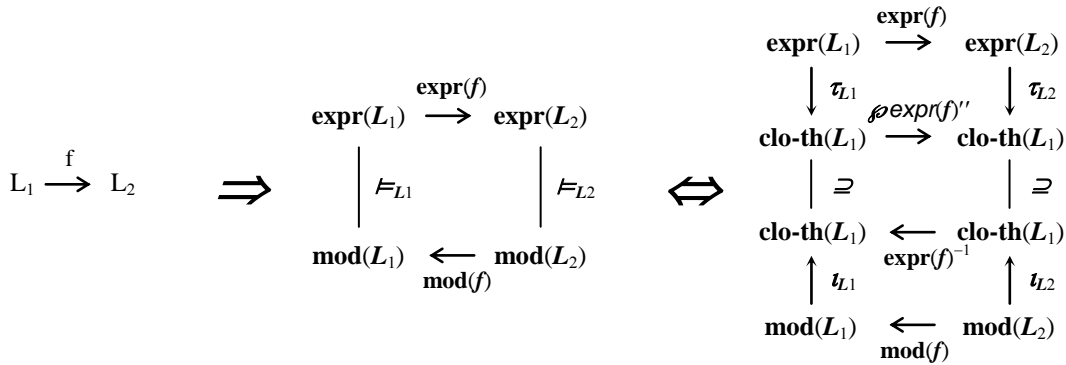

1st-order type language morphism  ⟹  *truth infomorphism*  ⟺  *truth concept morphism* = "lattice of theories passage"

## Fibered or Parametric Structure

It is important to note that we are dealing with a fibered or parametric structure here. The IFF truth concept lattice represents John Sowa's potentially infinite open-ended lattice of theories. However, there is no single lattice here but an infinite collection of lattices, where each truth classification "*truth-classification*($L$)" and each truth concept lattice "*truth-concept-lattice*($L$)" is based upon (indexed by) a particular 1st-order language $L$. So a particular object level ontology – be it a domain ontology, a middle level ontology, or an upper level ontology – will be an element of a lattice based on its 1st-order language. Internally, the various ontologies (theories) are connected by generalization/specialization and sideways jumping. Externally, the lattices in the IFF framework are connected in many different ways, such as by the FCA notions of apposition and subposition, etc. However, a very special and especially interesting connection is through 1st-order interpretations.

## The Category of Theories

The lattice of theories can serve as an introduction to the category of theories. Internally, each truth concept lattice links (orders) its theories by reverse set inclusion of their closures. Externally, concept lattices are linked by concept morphisms. More precisely, truth concept lattices are situated in the cocomplete quasicategory of large concept lattices and concept morphisms. The colimit operation, which fuses diagrams of concept lattices and concept morphisms into a single concept lattice, can perhaps profitably be thought of as an internalization operator.

The category of type languages Language has 1st-order type language as objects and type language morphisms as morphisms. Theories add axioms to languages. The category of theories Theory, which has 1st-order theories as objects and theory morphisms as morphisms, is compatible with the "lattices of theories" construction. The category of theories is indexed by the category of languages via the base language passage *lang* : Theory → Language, which maps a theory to its underlying type language. These contexts are illustrated in Figure 3, where the category of 1st-order theories is represented by the boldly enclosed upper rectangle, the category of closed theories Closed Theory is a sub-rectangle, and the category of 1st-order type languages Language is represented by the boldly enclosed lower rectangle. In Figure 3, each point in the rectangle representing the category of theories corresponds to a 1st-order theory, and each point in the rectangle representing the category of languages corresponds to a 1st-order type language. Two theories can be linked by a theory morphism, as illustrated by the theory morphism $f : T_1 \rightarrow T_2$ in the upper left part of the category of theories in Figure 3. The closure construction *clo* : Theory → Closed Theory maps the category of theories onto the subcategory of closed theories. This is illustrated by the association of the theory $T_0$ to its closure $clo(T_0)$ on the left side of the category of theories. In Figure 3, each small colored rectangle of the category of theories represents a fiber of theories (and theory morphisms) over a fixed type language. For example, the pea-green colored small rectangle in the lower middle part of the category of theories is the fiber over the language $L$. Since theory morphisms within a fiber are mapped to the identity type language morphism at the indexing language, they are equivalent to the (opposite) lattice ordering. This is illustrated by the theory order $\check{T}_1 \geq \check{T}_2$ in the pea-green colored small rectangle. The restriction of a fiber to the sub-category of closed theories corresponds precisely to the lattice of theories over the indexing type language. Sums in the category of theories correspond to meets in the lattices of theories (fibers). However, fusion and quotienting in the category of theories have no correspondents in the lattice of theories. The category of theories is not only compatible with the lattice of theories, but is also a proper extension of it.

---

[1] For a type language interpretation $h : L_1 \rightarrow expr(L_2)$, replace *expr*($f$) with the extension $h^{\#} : expr(L_1) \rightarrow expr(L_2)$.

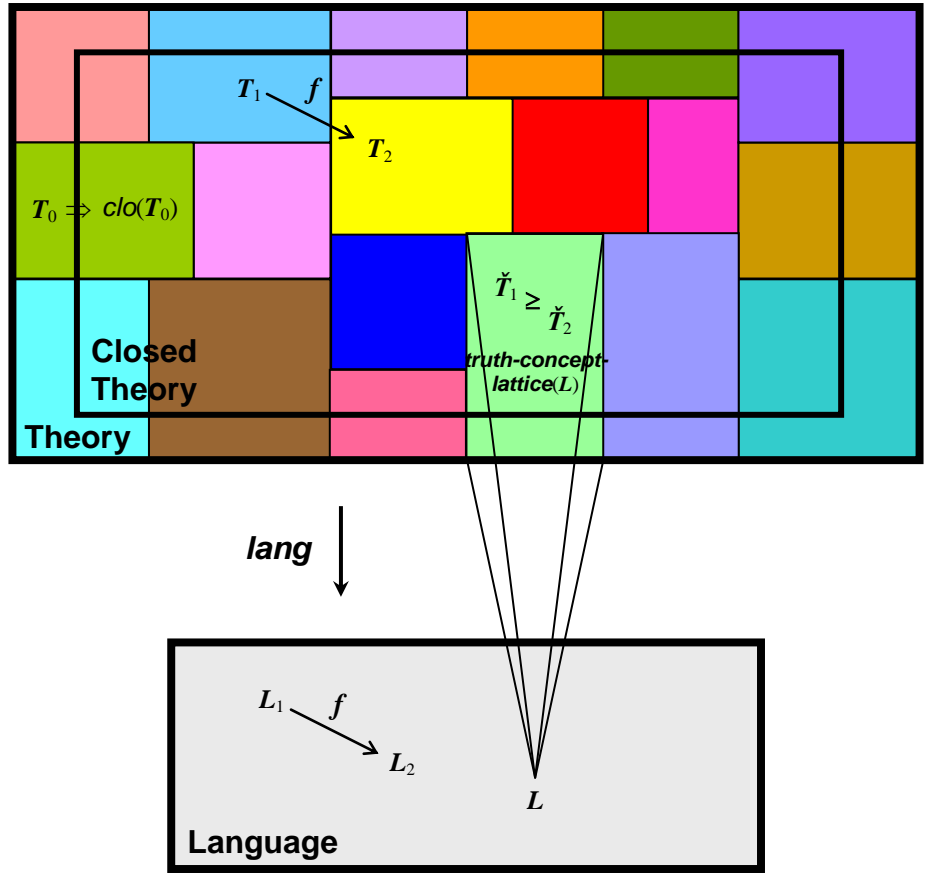

**Figure 3: Compatibility of the Category of Theories with the Lattice of Theories**

Analogy is represented in the IFF by type language and theory morphisms. A type language morphism

$f : L \to Ł$

from type language $L$ to type language $Ł$ maps (renames) the entity types, relation types and constants using the entity, relation and constant functions

- the *entity* type function $ent(f) : ent(L) \to ent(Ł)$,
- the *relation* type function $rel(f) : rel(L) \to rel(Ł)$,
- the *constant* function $const(f) : const(L) \to const(Ł)$,

respectively. Associated with a type language morphism $f : L \to Ł$ is the truth concept lattice of theories morphism

*truth-concept-morphism*$(f)$ : *truth-concept-lattice*$(L) \rightleftarrows$ *truth-concept-lattice*$(Ł)$

from the truth concept lattice of theories over $L$ to the truth concept lattice of theories over $Ł$.

## Interpretations

Type language interpretations extend the notion of type language morphisms. The notion of a $1^{st}$-order interpretation was discussed as example 4.11 on page 74 of the book by Barwise and Seligman. $1^{st}$-order interpretations are axiomatized in the IFF Type Language Namespace, a module in the IFF lower metalevel. A $1^{st}$-*order type language interpretation*

$$h : L_1 \to L_2$$

from a $1^{st}$-order type language $L_1$ to a $1^{st}$-order type language $L_2$ is an arity-preserving map from the relation types of $L_1$ to the expressions (formulas) of $L_2$. The category Theory$_{expr}$ has $1^{st}$-order type languages as objects and $1^{st}$-order type language interpretations as morphisms. There are two relevant passages:

$1^{st}$-order interpretation
⇒ infomorphism between truth classifications
⇒ concept morphism (adjoint pair of monotonic functions) between truth concept lattices

Each 1st-order interpretation $h : L_1 \to L_2$ defines a "truth infomorphism" between the associated truth classifications,

*truth-infomorphism*($h$) : *truth-classification*($L_1$) ⇌ *truth-classification*($L_2$)

This infomorphism defines a "truth concept morphism" (axiomatized in the truth namespace) containing an adjoint pair of monotonic functions between the associated truth concept lattices,

*truth-concept-morphism*($h$) : *truth-concept-lattice*($L_1$) ⇌ *truth-concept-lattice*($L_2$).

These monotonic functions map between the axiomatic truth concepts (the theories representing object-level ontologies) in a very semantic way.

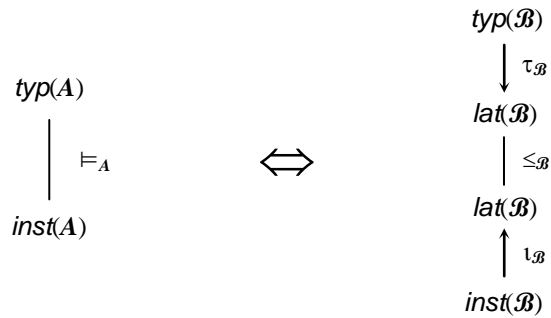

**Figure 1a: Classification *A***  **Figure 1b: Concept Lattice *B***

## *References*

## The Information Flow Framework

- Kent, Robert E. 2000. The Information Flow Foundation for Conceptual Knowledge Organization. In: *Dynamism and Stability in Knowledge Organization. Proceedings of the Sixth International ISKO Conference. Advances in Knowledge Organization* 7 (2000), pp. 111-117. Ergon Verlag, Würzburg.

- Kent, Robert E. 2002a. Distributed Conceptual Structures. In: *Proceedings of the 6th International Workshop on Relational Methods in Computer Science (RelMiCS 6). Lecture Notes in Computer Science* 2561. Springer, Berlin.

- Kent, Robert E. 2003. <u>The IFF Foundation for Ontological Knowledge Organization</u> In: *Knowledge Organization and Classification in International Information Retrieval. Cataloging and Classification Quarterly*. The Haworth Press Inc., Binghamton, New York.

## Channel Theory and Semantic Integration

- Y. Kalfoglou, M. Schorlemmer. <u>IF-Map: An Ontology-Mapping Method based on Information-Flow Theory</u>. *Journal of Data Semantics I*, LNCS 2800, Springer 2003.
- R. Kent. <u>The IFF Approach to Semantic Integration</u>. Presentation at the Boeing Mini-Workshop on Semantic Integration, 7 November 2002.
- R. Kent. <u>Semantic Integration in the IFF</u>. *ISWC'03 Semantic Integration Workshop*. Sanibel Island, Florida, USA, 2003.
- R. Kent. <u>Semantic Integration in the IFF</u>. Semantic Interoperability and Integration. Dagstuhl Seminar 04391, September 2004.
- M. Schorlemmer, Y. Kalfoglou. <u>Using Information-Flow Theory to Enable Semantic Interoperability</u>. In *Artificial Intelligence Research and Development*, volume 100 of Frontiers of Artificial Intelligence and Applications. IOS Press, 2003.
- M. Schorlemmer, Y. Kalfoglou. <u>On Semantic Interoperability and the Flow of Information</u>. *ISWC'03 Semantic Integration Workshop*. Sanibel Island, Florida, USA, 2003.

## Mathematical and Philosophical Foundations

- Barwise, Jon and Seligman, Jerry. 1997. <u>Information Flow: The Logic of Distributed Systems</u>. Cambridge Tracts in Theoretical Computer Science 44. Cambridge University Press.
- Chang, C. C. and Keisler, H. J. 1973. <u>Model Theory</u>. Studies in Logic and the Foundations of Mathematics 73. Amsterdam: North Holland.
- Mac Lane, Saunders. 1971. <u>Categories for the Working Mathematician</u>. New York/Heidelberg/Berlin: Springer-Verlag. New edition (1998).
- Pitts, Andrew M. 2000. <u>Categorical Logic</u>. In S. Abramsky, D.M. Gabbay, and T.S.E. Maibaum, editors, Handbook of Logic in Computer Science, Volume 5: Algebraic and Logical Structures. Chapter 2. Oxford University Press.
- Seely, Robert A.G. <u>Hyperdoctrines, natural deduction, and the Beck condition</u>. Zeitschrift f. math. Logik und Grundlagen d. Math. 29 (1983) 505-542.
- Sowa, John S. 2000. <u>Knowledge Representation: Logical, Philosophical, and Computational Foundations</u>. Brookes/Cole: Pacific Grove, CA, USA.

### i Common Notions

The notion of *classification* is central to IF. We identify this with the notion of *formal context* in FCA. A *classification* (Figure 1a)

$$A = \langle \mathit{inst}(A), \mathit{typ}(A), \vDash_A \rangle$$

is identical to a binary relation. However, from a category-theoretic standpoint, the category of classifications is very different from the category of relations, since their morphisms are very different. A classification consists of a collection of *instances* $\mathit{inst}(A)$ identified with the first or source collection of a binary relation, a collection of *types* $\mathit{typ}(A)$ identified with the second or target collection of a binary relation, and a binary *classification relation* $\vDash_A$ identified with the extent collection of a binary relation. A classification is small when all collections are sets. A classification is large when all collections are classes. In FCA, instances are called (*formal*) *objects*, types are called (*formal*) *attributes*, and the classification relation is called the *incidence relation* between object and attributes.

### ii Concept Lattice

The notion of a concept lattice is abstractly defined in the paper "Distributed Conceptual Structures" by Kent. See the IFF references. An (abstract) *concept lattice* (Figure 1b) ($\mathcal{B}$ for "Begriffsverbande" meaning concept lattice in German)

$$\mathcal{B} = \langle \mathit{lat}(\mathcal{B}), \mathit{inst}(\mathcal{B}), \mathit{typ}(\mathcal{B}), \iota_\mathcal{B}, \tau_\mathcal{B} \rangle$$

consists of a complete lattice $\mathit{lat}(\mathcal{B})$, two collections $\mathit{inst}(\mathcal{B})$ and $\mathit{typ}(\mathcal{B})$ called the instance collections and the type collections of $\mathcal{B}$, respectively; along with two functions, an instance embedding function

$$\iota_\mathcal{B} : \mathit{inst}(\mathcal{B}) \to \mathit{lat}(\mathcal{B})$$

and a type embedding function

$$\tau_\mathcal{B} : \mathit{typ}(\mathcal{B}) \to \mathit{lat}(\mathcal{B}),$$

which satisfying the following conditions.

The image $\iota_\mathcal{B}(\mathit{inst}(\mathcal{B}))$ is join-dense in $\mathit{lat}(\mathcal{B})$.
The image $\tau_\mathcal{B}(\mathit{typ}(\mathcal{B}))$ is meet-dense in $\mathit{lat}(\mathcal{B})$.

The double arrow in the center of Figure 1 illustrates the intuitive and operative equivalence between classifications and concept lattices. This is known as the basic theorem of FCA. More abstractly, this is also a categorical equivalence. As the FCA shows, any concept lattice has an underlying classification – just compose the three relations in Figure 1b:

$$i \vDash_\mathcal{B} t \text{ iff } \iota_\mathcal{B}(i) \leq_\mathcal{B} \tau_\mathcal{B}(t).$$

### iii Formal Concept

A *formal concept* $C$ of a classification $A$ is a pair of subcollections $C = (X, Y)$, where the subcollection of instances $X \subseteq \mathit{inst}(A)$ called the *extent* of $C$ and the subcollection of types $Y \subseteq \mathit{typ}(A)$ called the *intent* of $C$ are related as follows:

$$X = Y' = \{i \in \mathit{inst}(A) \mid i \vDash_A t \text{ for all } t \in Y\}, \text{ and}$$

$$Y = X' = \{t \in \mathit{typ}(A) \mid i \vDash_A t \text{ for all } i \in X\}.$$

It is important to note that a formal concept $C$ is determined by either its extent $X$ (since $Y = X'$) or its intent $Y$ (since $X = Y'$); in fact as we discuss below the IFF axiomatization for the lattice of theories uses the intent.

If $C_1 = (X_1, Y_1)$ and $C_2 = (X_2, Y_2)$ are two formal concepts of a classification, $(X_1, Y_1)$ is called a *subconcept* of $(X_2, Y_2)$, denoted by $C_1 \leq_A C_2$, when $X_1 \subseteq X_2$ or equivalently $Y_1 \supseteq Y_2$. In this case, $C_2$ is called a *superconcept* of $C_1$. The relation $\leq_A$ is called the *concept order* of the classification $A$. The collection $\mathcal{B}(A)$ of all formal concepts of $A$ with this order is called the *concept lattice* of the classification $A$. This is a complete lattice in which the infimum and supremum of a subcollection of formal concepts $\mathcal{C} = \{C_n = (X_n, Y_n) \mid n \in N\}$ are given by:

$$\sqcap_A \mathcal{C} = \sqcap_A \{C_n \mid n \in N\} = (\cap_{n \in N} X_n, (\cup_{n \in N} Y_n)''), \text{ and}$$

$$\sqcup_A \mathcal{C} = \sqcup_A \{C_n \mid n \in N\} = ((\cup_{n \in N} X_n)'', \cap_{n \in N} Y_n).$$

In addition, there is an instance embedding function

$$\iota_A : \mathit{inst}(A) \to \mathcal{B}(A)$$

defined by $\iota_A(a) = (\{a\}'', \{a\}')$ for any instance $a \in \mathit{inst}(A)$, and there is a type embedding function

$$\tau_A : \mathit{typ}(A) \to \mathcal{B}(A)$$

define d by $\tau_A(\alpha) = (\{\alpha\}', \{\alpha\}'')$ for any type $\alpha \in \mathit{typ}(A)$. These satisfy the conditions: the image $\iota_A(\mathit{inst}(A))$ is join-dense in $\mathcal{B}(A)$, and the image $\tau_A(\mathit{typ}(A))$ is meet-dense in $\mathcal{B}(A)$.